\documentclass[twocolumn,aps,prl,showpacs,amsmath,superscriptaddress]{revtex4}
\usepackage{amssymb}
\usepackage{mathrsfs}
\usepackage{graphicx}
\usepackage{float}
\usepackage{txfonts}
\usepackage[normalem]{ulem}
\usepackage{color}

\begin{document}

\title{Chiral topological insulating phases from three-dimensional nodal loop semimetals}

\author{Linhu Li}
\affiliation{Beijing Computational Science Research Center, Beijing 100089, China}
\affiliation{CeFEMA, Instituto Superior
T\'ecnico, Universidade de Lisboa, Av. Rovisco Pais, 1049-001 Lisboa,  Portugal}
\author{Chuanhao Yin}
\affiliation{Beijing National Laboratory for Condensed Matter Physics,
Institute of Physics, Chinese Academy of Sciences, Beijing 100190, China}
\author{Shu Chen}
\affiliation{Beijing National Laboratory for Condensed Matter Physics,
Institute of Physics, Chinese Academy of Sciences, Beijing 100190, China}
\affiliation{School of Physical Sciences, University of Chinese Academy of Sciences, Beijing, 100049, China}
\affiliation{Collaborative Innovation Center of Quantum Matter, Beijing, China}
\author{Miguel A. N. Ara\'ujo}
\affiliation{Beijing Computational Science Research Center, Beijing 100089, China}
\affiliation{CeFEMA, Instituto Superior
T\'ecnico, Universidade de Lisboa, Av. Rovisco Pais, 1049-001 Lisboa,  Portugal}
\affiliation{Departamento de F\'{\i}sica,  Universidade de \'Evora, P-7000-671,
\'Evora, Portugal}

\begin{abstract}

We identify a topological $\mathbb{Z}$ index for three dimensional chiral insulators with
$P*T$ symmetry where two Hamiltonian terms define a nodal loop.
Such systems may belong in the AIII or DIII symmetry class.
%We study the chiral topological insulating phases in
%three-dimensional $P*T$ nodal loop semimetals with anticommuting gap terms,
%and the topological properties can be characterized by a $Z$ invariant of
The  $\mathbb{Z}$ invariant is a winding number assigned to the nodal loop
and  has a correspondence to the geometric relation between the nodal loop
and the zeroes of the gap terms.
Dirac cone edge  states under open boundary conditions
%have also been investigated, which hold Dirac cones with the
are in correspondence with the winding numbers assigned to the nodal loops.
We verify our method with the low-energy effective Hamiltonian
of a three-dimensional material of topological insulators in the Bi$_2$Te$_3$ family.

\end{abstract}

\maketitle
\emph{Introduction.-}
%The research on topological phases is one of the most intriguing aspects
%in condensed matter physics\cite{Topological1,Topological2,classes}.
Topological insulators (TIs) in three dimensions (3D)
having time reversal symmetry
can be characterized by  $\mathbb{Z}_2$ numbers defined on some discrete
momenta\cite{Topological1,Topological2,classes,3DTI_1,3DTI_2},
which is equivalent to a quantized invariant expressed as an integral
over the entire Brillouin Zone (BZ)\cite{3DTI_3}.
There are then two types of TI's, strong and weak,
according to whether there is an
odd or even number of Dirac cone surface states,
respectively\cite{3DTI_2}.
Besides the time reversal TIs, there is a class of chiral TIs which are described by a
$\mathbb{Z}$-type topological invariant\cite{cTI}.
The geometrical representation of a
topological invariant in some vector spaces provides an
intuitive way to analyze the topological nature of
many systems\cite{Geometry1,Geometry2,Geometry3,Geometry4,Geometry5}.

On the other hand, a transition point between topologically
different insulating phases can be viewed as a semimetal phase with nontrivial
topology in its gap closing points\cite{transition1,transition2}.
%The research on topological phases is becoming one of the most intriguing aspects in condensed matter physics. Topological phases are classified according to quantized topological invariants rather than some continuous order parameters\cite{Topological1,Topological2}. Among them are the widely studied topological gapped systems, i.e. topological insulators (TI) and superconductors (TSC), which can be classified into ten topological classes by their time-reversal, particle-hole, and chiral symmetries, and five of which can support topologically nontrivial phases depending on the dimension of the systems\cite{classes}. These systems hold in-gap surface or edge states under open boundary condition, which are robust metallic modes on the surface for TIs, or the quasi-particle modes of Majorana fermions in TSCs.
%Gapless systems have also been found to support topological phases, known as
Topological semimetals (TSMs)\cite{TSM1,TSM2,TSM3,TSM4,TSM5,TSM6,TSM7}
%Compared to normal metals, TSMs
have a Fermi surface (FS) with reduced dimension.
While a 3D normal metal has a two-dimensional (2D) FS,  a TSM has a one-dimensional
(1D) or zero-dimensional (0D) FS at half-filling.
3D systems with 0D FS are known as the Weyl\cite{TSM1} or Dirac\cite{TSM2} semimetals.
In these systems, the two bands touch linearly at discrete gap closing points
 in the BZ, and
hold topologically protected edge states under open boundary conditions
(OPC), such as for instance, the Fermi arcs.
More recently,
3D nodal line semimetals\cite{TNL0,TNL1,TNL2,TNL3,TNL_Fang1,TNL_Fang2}
have attracted growing attention.
In such systems, the linear band touching points  form one or several 1D lines in the BZ.

One of the most interesting cases is when the nodal lines form closed, nodal loops (NLs).
A NL can be classified in either of two classes,
according to whether it carries a $\mathbb{Z}_2$ monopole charge or not.
The one without a monopole charge can shrink into a point and disappear,
and is topologically trivial in this sense.
%These two types of
NLs are protected by the
combination of inversion and time reversal symmetries,
$P*T$, for spinless systems, while additional symmetries are required to protect
NLs in 3D systems with spin-orbital coupling\cite{TNL_Fang1,TNL_Fang2}.
On the other hand, NL semimetals have also been
%proposed  and
studied in 2D\cite{TNL_2D,TNL_Linhu}.
In this case, although the NL itself does not carry topological charge,
the addition of  some chiral gap terms can make the system
become topological and insulating, where the topological invariant
is given by a winding number defined along the NL\cite{TNL_Linhu}.
An interesting question to address is,
what effect can gap terms have on a 3D NL semimetal?

In this paper, we study a spin-1/2 system with $P*T$ symmetry,
and show that anticommuting mass gap terms can drive a 3D NL semimetal into a chiral TI,
which can be characterized by a integer winding number defined along each NL.
This winding number is determined by the geometric relation between the NL and the
zeroes of the  gap terms.
Although the gap terms may be initially considered small,
our results only depend on their zeroes, so that {\it it holds valid for any finite terms
that gap out the NLs.}
The system's surface states may hold an odd or even number of Dirac cones,
and their existence has a correspondence to the NL winding number.
In this sense, the NL can serves as an indicator of the topological properties of a 3D insulator.
In order to show the utility of our theory, we apply it to a 3D material of
the topological Bi$_2$Te$_3$ family and give a brief discussion.

\emph{Minimal model.-}
We begin our discussion with a simplest two band model for $P*T$ symmetry-protected NL semimetals:
\begin{eqnarray}
H_0=(m-k_x^2-k_y^2)\sigma_x+k_z\sigma_z,
\label{hzero}
\end{eqnarray}
with $\sigma$ the Pauli matrices acting on an orbital space.
%As the parity and time-reversal symmetry require $k\rightarrow-k$ and $k\rightarrow-k,~i\rightarrow -i$ respectively,
For spinless system, the $P*T$ symmetry is simply given by
the  complex conjugation and a unitary matrix, $P$,  such that
the Hamiltonian satisfies $PH_0^*(k)P^{-1}=H_0(k)$. In this case $P=1$ and
%\textbf{the $P*T$ simply transforms $H(k)$ into $H^*(k)$ for spinless systems.}
the $P*T$ symmetry ensures the absence of the second Pauli matrix.
The nodes of $H_0$ yield a 1D solution,
a NL with $k_z=0$ and $k_x^2+k_y^2=m$.
Introducing a $\sigma_y$ term, $H=H_0+h_y\sigma_y$, not only breaks
$P*T$ symmetry but $h_y$ also
%This symmetry can be broken by introducing a $\sigma_y$ term,
%\begin{eqnarray}
%H=H_0+h_y\sigma_y.
%\end{eqnarray}
%This
serves as an effective mass term, which can be either $k$-independent or related with $k$.
In the former case, it opens a stable gap in the BZ, which drives the system into a trivial
insulator.  If $h_y$ is a function of $k$,
%$h_y=0$ shall give a 2D solution,
%which will cut the NL in
the nodes of $H$ may
be pairs of points, and the system is a Weyl semimetal.

We extend this model by including the spin degree of freedom
and write the Hamiltonian with  $P*T$ symmetry as
\begin{eqnarray}
H=H_0+h_{yx}\sigma_y s_x+h_{yy}\sigma_y s_y+h_{yz}\sigma_y s_z,
\label{hfive}
\end{eqnarray}
where the Pauli matrices $s_j$ act in spin space.
 $P*T$ symmetry now reads
$Ps_yH_0^*(k)s_yP^{-1}=H_0(k)$, with $P=1$ and satisfies $(P*T)^2=-1$,
%\textbf{and the $P*T$ transformation acts as $H(k)\rightarrow s_yH^*(k)s_y$ for spin-1/2 systems.}
Here we use the labels $h_{ij}$ to represent the term of $\sigma_i$ and $s_j$,
with $i=0$ or $j=0$ for the identity matrix in the corresponding subspace.
These five terms form an anticommuting set of Dirac matrices,
but we note that there are also other equivalent choices\cite{Arfken}.
%This Hamiltonian has two doublet bands, and the
The spectrum of (\ref{hzero})-(\ref{hfive})
is simply given by
\begin{eqnarray}
E_{\pm}=\pm\sqrt{h_{x0}^2+h_{z0}^2+h_{yx}^2+h_{yy}^2+h_{yz}^2}.
\end{eqnarray}
%Comparing to the spinless case, the
The effective mass gapping out the NL is now  $|\mathbf{h}_y|$, where
%\begin{eqnarray}
%h_y=|\mathbf{h}_y|,
%\end{eqnarray}
%with
$\mathbf{h}_y=(h_{yx},h_{yy},h_{yz})$.
Requiring $\mathbf{h}_y=0$ may give a solution of points (0D),
lines (1D) or surfaces (2D),
depending on the number of non-zero $k$-dependent terms it has.
%If $|\mathbf{h}_y|$ contains only one non-zero $k$-dependent term,
%$h_y=0$ shall give a 2D solution, and the situation is the same as the spinless one
%discussed above.

\begin{figure}
\includegraphics[width=0.8\linewidth]{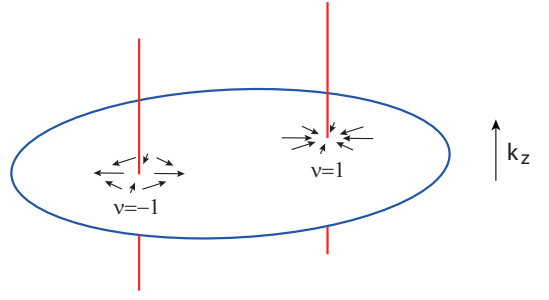}
\caption{A sketch of the winding number of a NL. The blue circle indicates the NL given by $H_0=0$,
the red lines indicate the 1D solution of $\mathbf{h}_y=0$ for models of the chiral class,
and each line gives a singularity in the integrand of (\ref{winding})
in the NL plane.
The arrows around each singularity show the direction of $\mathbf{h}_y$ near this point,
and the corresponding winding number of these points are labeled in the sketch.
The winding number of the NL, $\nu_{NL}$, is given by the summation of winding numbers
of every singularity within the NL, as the integral path can be smoothly transformed into
circles around each singularity.}\label{fig1}
\end{figure}

If $|\mathbf{h}_y|$ contains two non-zero $k$-dependent terms,
$\mathbf{h}_y=0$ shall give one or several 1D lines. Thus the system is generally an insulator,
as the gap closing condition requires the crossing of the NL and these 1D lines,
which is accidental.
Such a four-component Dirac Hamiltonian describes a chiral topological insulator\cite{cTI},
as the model satisfies $SH(k)S^{-1}=-H(k)$,
with the chiral operator $S$ given by the absent fifth Dirac matrix.
In the absence of time reversal symmetry, the system belongs to the AIII class
and can be characterized by a Z invariant\cite{classes}.
We define a winding number of $\mathbf{h}_y$ along the NL\cite{TNL_Linhu},
\begin{eqnarray}
\nu_{NL}=\oint_{NL}\frac{h_2 dh _1-h_1 dh _2}{|\mathbf{h}_y|^2},\label{winding}
\end{eqnarray}
can be shown to be
equivalent to a Berry phase of the occupied Bloch bands at half filling\cite{Berry}.
Here, $h_1$ and $h_2$ denote the two $k$-dependent terms of $\mathbf{h}_y$.
If we consider the $k_x-k_y$ plane that contains the NL,
the intersection of the 1D lines and the plane produces a series of singularities in the plane,
and the winding number (\ref{winding}) of the NL is simply the summation of the windings around
the singularities within the NL, as shown in Fig.\ref{fig1}.
This winding number may take on any integer value,
as it is only associated with the number of lines going through the NL.
We also note that in the presence of time reversal symmetry,
this model would fall into the CII class, which is described by a $Z_2$
topological index instead\cite{classes}.

Finally, if all the three terms of $\mathbf{h}_y$ are depended on $k$,
$h_y=0$ shall give one or several 0D points.
In this case, the intersection of these points and the plane of the NL is also accidental.
From the symmetry classification point of view,
the presence of the fifth Dirac matrix breaks the chiral symmetry,
and the model falls into the  A class, which is non-topological in 3D.
In other words, we could smoothly move a singularity out of the NL without closing the gap.

%\section{lattice models}
\emph{Winding numbers and geometry of the loops for a lattice model.-}
In order to reveal the topological properties described by the NL winding number, we next consider a lattice model described by an anticommuting set of Dirac matrices $\mathbf{\Gamma}=(\sigma_x s_0,\sigma_z s_0,\sigma_y s_x,\sigma_y s_y,\sigma_y s_z)$, as
\begin{eqnarray}
H&=&\mathbf{h}(k)\cdot\mathbf{\Gamma},\label{total_H}
%H=h_{x0}\sigma_x s_0+h_{z0}\sigma_z s_0+h_{yx}\sigma_y s_x+h_{yy}\sigma_y s_y+h_{yz}\sigma_y s_z,\label{total_H}
\end{eqnarray}
with
\begin{eqnarray}
\mathbf{h}(k)&=&(h_{x0},h_{z0},h_{yx},h_{yy},h_{yz}),\nonumber\\
h_{x0}&=&\mu-t_{\parallel}(\cos{k_x}+\cos{k_y}),\nonumber\\
h_{z0}&=&-t_{\perp}\cos{k_z},\nonumber
\end{eqnarray}
which form two NLs in $k_x-k_y$ plane for $k_z=\pi/2$ and $-\pi/2$ when $0<|\mu/t_{\parallel}|<2$.
The position and shape of the NLs are only associated with the ratio of
$\mu$ and $t_{\parallel}$, hence we can choose $t_{\parallel}=t_{\perp}=1$ for
the sake of simplicity.
In the following we only consider the case with positive $\mu$, with the center of the NLs
given by $k_x=k_y=0$. For negative $\mu$, the center of the NLs is at $k_x=k_y=\pi$,
and a similar discussion applies.
The two NLs of $k_z=\pm\pi/2$ give two independent winding numbers, $\nu_{NL}^{\pm}$,
respectively, and we define the total winding number of the system as
\begin{eqnarray}
\nu_{sum}=\nu_{NL}^+ +\nu_{NL}^-.
\end{eqnarray}
Without loss of generality, here we choose $h_{yx}=0$ to preserve a chiral symmetry
with the operator $S=\sigma_y s_x$. The other two gap terms of $\mathbf{h}_y$ are functions of $k$,
and $\mathbf{h}_y=0$ gives 1D lines in the BZ.
We consider the following form of $\mathbf{h}_y$ that breaks time reversal symmetry:
\begin{eqnarray}
h_{yy}&=&-t_y\sin{k_y}\\
h_{yz}&=&\mu_s-t_x\sin{k_x}-t_z\sin{k_z},
\end{eqnarray}
and the system falls into the AIII class.
$|\mathbf{h}_y|=0$ gives some 1D lines in the $k_x-k_z$ plane with
$k_y=0$ or $\pi$, which may or may not be enclosed by the NLs.
%Here a finite $t_y$ is necessary to give a 1D solution, but the exact value is irrelevant as it is the only term in $h_{yy}$. As for the rest three parameters, only two of them are independent.
We would also like to point out that although the gap terms need be small
for the system to preserve a NL like structure,
{\it the topological properties are not related to the exact value of these terms,
but only to the ratios between them}.
%On the other hand, for the specific case we choose, $\mathbf{h}_y=0$ will give a 2D solution if $t_y=0$, or 1D lines parallel to the NLs if $t_x=0$.
For the sake of simplicity, we choose $t_y=t_x=1$ and positive $\mu_s$ hereafter.

\begin{figure}
\includegraphics[width=0.8\linewidth]{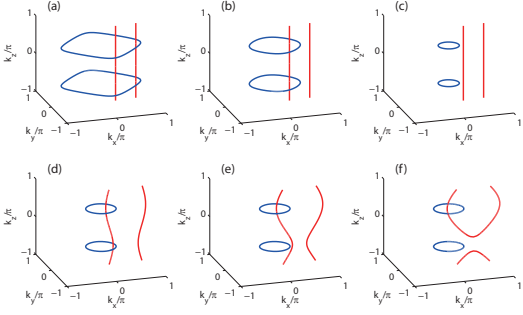}
\caption{The NLs and the 1D lines given by $|\mathbf{h}_y|=0$,
represented by blue and red lines respectively.
For the latter, we only show the lines with $k_y=0$,
as the ones for $k_y=\pi$ lie outside the NLs.
(a) $\mu=0.2$, $\mu_s=0.8$ and $t_z=0$; (b) $\mu=1$, $\mu_s=0.8$ and $t_z=0$; (c) $\mu=1.8$, $\mu_s=0.8$ and $t_z=0$; (d) $\mu=1.6$, $\mu_s=0.5$ and $t_z=0.2$; (e) $\mu=1.6$, $\mu_s=0.5$ and $t_z=0.4$; (f) $\mu=1.6$, $\mu_s=0.5$ and $t_z=0.6$. }\label{fig2}
\end{figure}

We first consider a simple case with $t_z=0$. In this case, the 1D solution of
$\mathbf{h}_y=0$ gives four lines perpendicular to $k_x-k_y$ plane,
two with $k_y=0$ and two with $k_y=\pi$. For positive $\mu$, the pair of lines with
$k_y=\pi$ are always outside the loop.
By tuning $\mu_s$ and $\mu$, the NLs may enclose
$2$, $1$ or $0$ lines with $k_y=0$, as shown by Fig.\ref{fig2}(a)-(c).
%, in the regions of $\mu-1<-\sqrt{1-\mu_s^2}$, $-\sqrt{1-\mu_s^2}<\mu-1<\sqrt{1-\mu_s^2}$ and $\mu-1>\sqrt{1-\mu_s^2}$ respectively
However, the windings of these lines in the $k_x-k_y$ plane have opposite values
(as in Fig.\ref{fig1}),
and the NL
enclosing either $0$ or $2$ lines will result in $\nu_{NL}=0$.
On the other hand,
as the system preserves a reflection symmetry along $z$ direction,
%}\sout{as the 1D solution is independent from $k_z$,}
each line will be enclosed by either two or none of the NLs,
hence the total winding number, $\nu_{sum}$, in this case is always even.

In the presence of a nonzero $t_z$, %\textbf{
the reflection symmetry is broken, and
%}
the lines of $|\mathbf{h}_y|=0$ will change shape with $t_z$ and eventually form a closed ring,
as shown in Fig.\ref{fig2}(d)-(f).
In this case, an enclosed line will cross one of the NLs at some point,  resulting
in a topological phase transition.
After this transition, the system has an odd winding number, $\nu_{sum}=1$,
as only one of the NLs encloses a singularity.

\emph{edgestates and phase diagram.-}
The topological properties of a 3D topological insulator can be represented by the number of
Dirac cones in the edge states under OBC.
Next, we apply the method in Ref.\cite{edge_state} to study the edge states in our model.
The existence of edge states and their eigenenergies,
under OBC,  are associated with the bulk topology
of the system, which can be seen by the trajectory of $\mathbf{h}(k)$ in the 5-component
vector space formed by the Dirac matrices $\mathbf{\Gamma}$.
Here we choose a surface plane perpendicular to the $x$ direction by fixing $k_y$ and $k_z$,
and study the corresponding edge states as an example.
Edge states in the other two directions can also be studied in this way,
and give similar results are obtained.
The Hamiltonian terms associated with $k_x$ give an elliptical
 trajectory of $\mathbf{h}$
in the 1-5 plane in $\mathbf{\Gamma}$ space,
\begin{eqnarray}
\mathbf{h}(k_x)=(-\cos{k_x},0,0,0,-\sin{k_x}).
\end{eqnarray}
The remaining Hamiltonian terms,
\begin{eqnarray}
\mathbf{h}^0=(\mu-\cos{k_y},-\cos{k_z},0,-\sin{k_y},\mu_s-t_z\sin{k_z}),
\end{eqnarray}
can be viewed as the vector from the origin of the vector space to the center of
the ellipse $\mathbf{h}(k_x)$. The parallel and perpendicular components of $\mathbf{h}^0$ to the 1-5 plane are given by
\begin{eqnarray}
\mathbf{h}^0_{\parallel}&=&(\mu-\cos{k_y},0,0,0,\mu_s-t_z\sin{k_z})\,,\\
\mathbf{h}^0_{\perp}&=&(0,-\cos{k_z},0,-\sin{k_y},0).
\end{eqnarray}
The existence of edge states depends on whether the ellipse
%projection of the  trajectory of
$\mathbf{h}(k_x)$ encloses the point $\mathbf{h}^0_{\parallel}$,
and this condition reads
\begin{eqnarray}
|\mathbf{h}^0_{\parallel}|=\sqrt{(\mu-\cos{k_y})^2+(\mu_s-t_z\sin{k_z})^2}<1\,.\label{condition}
\end{eqnarray}
Provided that Eq.(\ref{condition}) holds, the edge state energies
are given by
%the distance between origin and the plane that contain $\mathbf{h}(k_x)$, as
\begin{eqnarray}
E_{\pm}=|\mathbf{h}^0_{\perp}|=\pm\sqrt{\cos^2{k_z}+\sin^2{k_y}}.
\end{eqnarray}

Candidate Dirac cones at  $(k_y,k_z)=(0,\pm\frac{\pi}{2})$
must satisfy the inequality (\ref{condition}).
When $t_z=0$, Eq.(\ref{condition}) becomes
\begin{eqnarray}
(\mu-1)^2<1-\mu_s^2\,,
\end{eqnarray}
for both $k_z=\pm\frac{\pi}{2}$. In other words, it gives either no Dirac cone, or a pair of
Dirac cones at  $(k_y,k_z)=(0,\pm\frac{\pi}{2})$.
For nonzero $t_z$, there may exist $0$, $1$ or $2$ Dirac cones depending on the parameters, as the condition (\ref{condition}) becomes
\begin{eqnarray}
(\mu_s\mp t_z)^2<2\mu-\mu^2\,,
\end{eqnarray}
for $k_z=\pm\frac{\pi}{2}$, respectively. In Fig.\ref{fig2} we display phase diagrams showing
the number of Dirac cone edge states. These results are also
in consistence with the winding numbers of the NLs, as shown in the figure.

\begin{figure}
\includegraphics[width=0.45\linewidth]{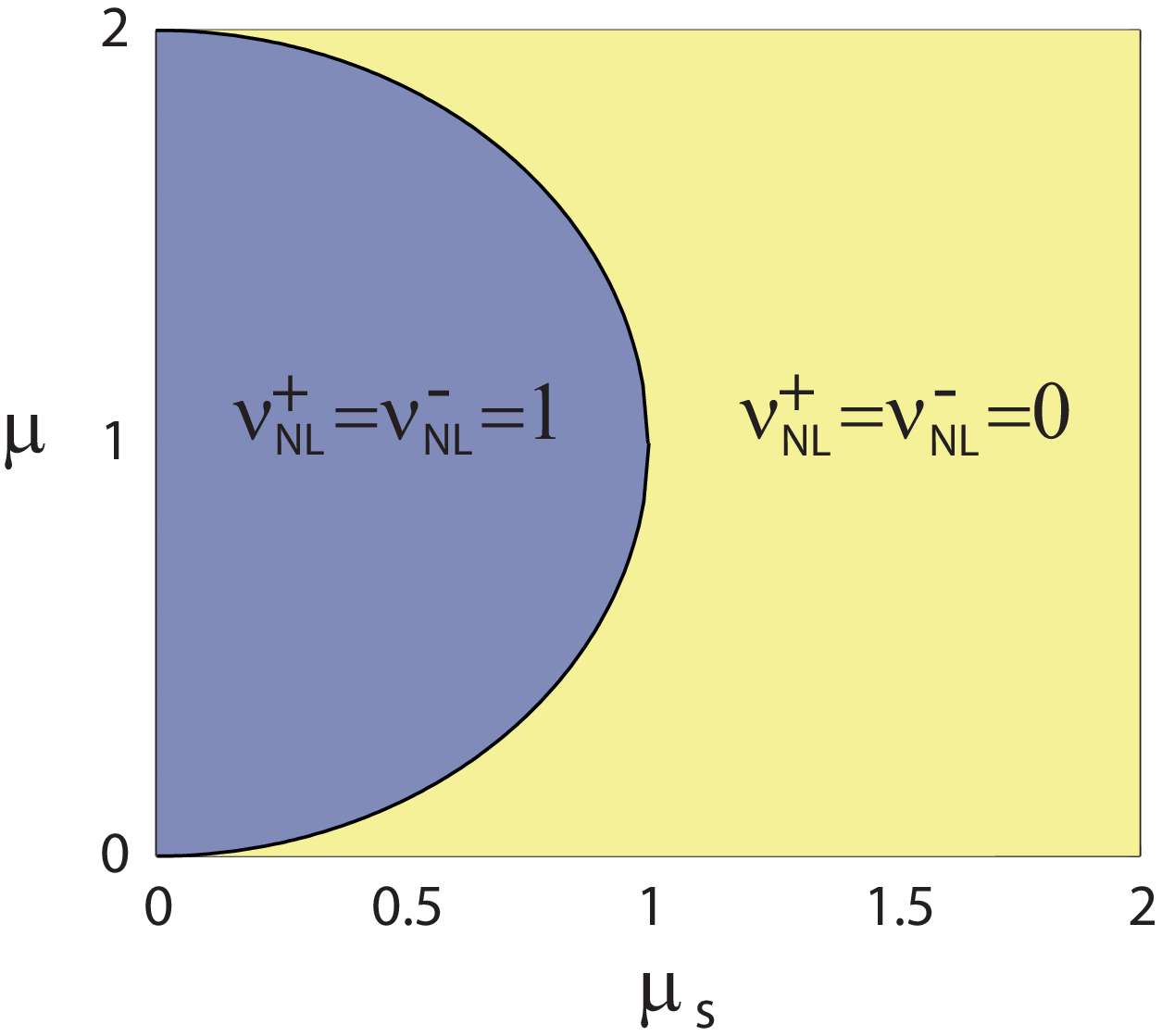}
\includegraphics[width=0.45\linewidth]{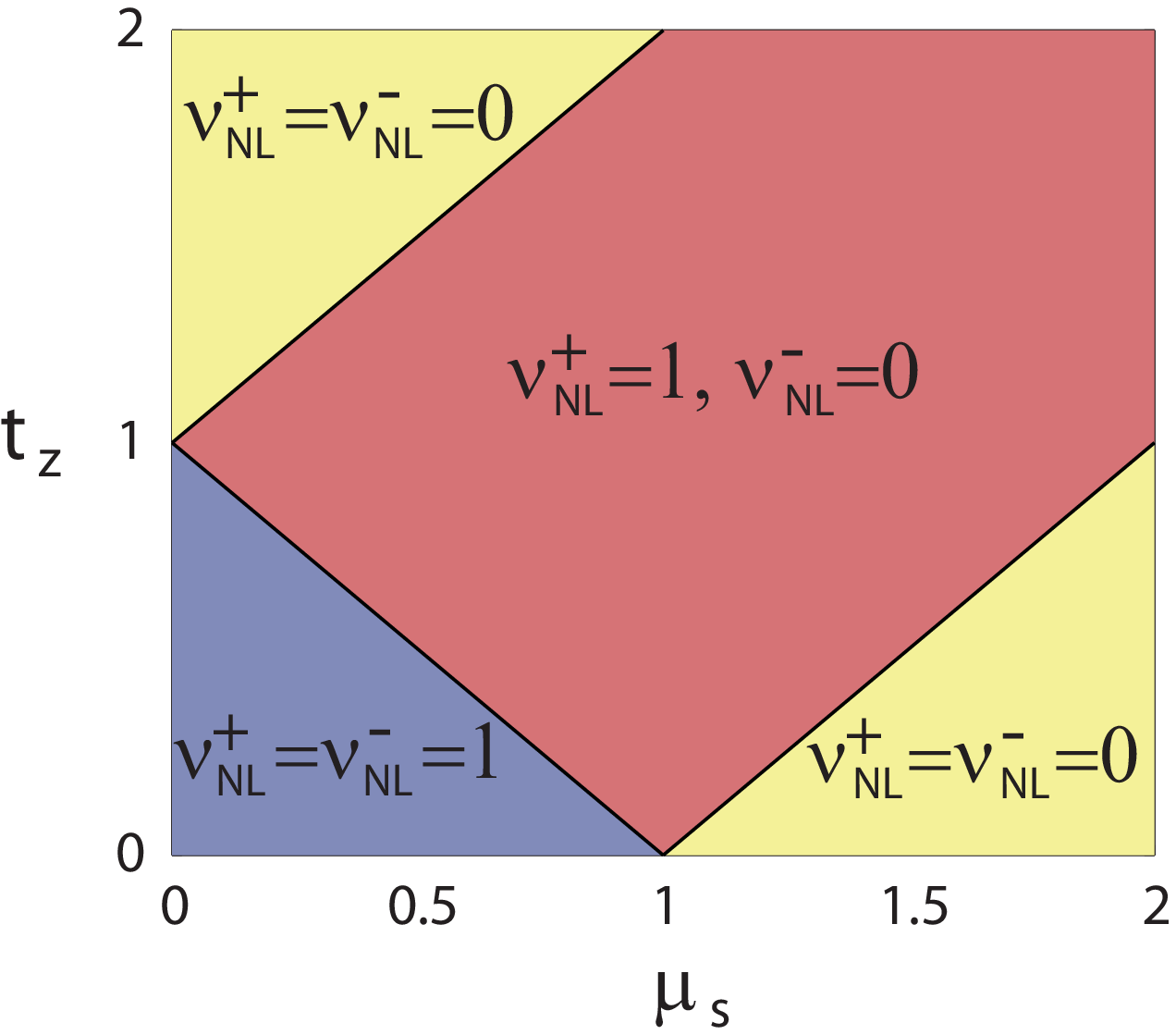}
\caption{Phase diagram with $t_z=0$ (left) and $\mu=1$ (right). The number of Dirac cones in edge states equals to the total winding number of the model, $\nu_{sum}=\nu_{NL}^+ +\nu_{NL}^-$.
In the left panel, the yellow region with $\nu_{NL}^+=\nu_{NL}^-=0$ includes three different situations: i) two lines are enclosed by the NLs when $\mu_s<1$ and $\mu<1$; ii) all the lines are out of the NLs when $\mu_s<1$ and $\mu>1$; and iii) not any line exsits when $\mu_s>1$.}\label{fig3}
\end{figure}

\begin{figure}
\includegraphics[width=1\linewidth]{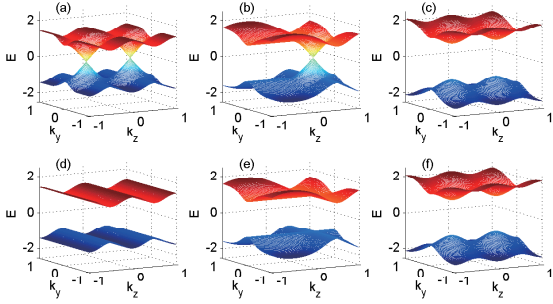}
\caption{The spectra of the two doublet bands nearest to $E=0$, (a)-(c) for open boundary condition along $x$, and (d)-(f) for period boundary condition. The three columns are for three topologically different phases, while the left two are topologically nontrivial. The parameters are $\mu=1$ and (a), (d) $\mu_s=t_z=0$; (b), (e) $\mu_s=t_z=1$; and (c), (f) $\mu_s=2$, $t_z=0$.}\label{fig4}
\end{figure}

In order to visualize the edge states, next we choose open boundary condition in $x$ direction and rewrite the Hamiltonian as
a tight-binding between planes
\begin{eqnarray}
H&=&\sum_{n}\hat{c}^{\dagger}_{n}U_d\hat{c}_{n}+\sum_{n}\frac{1}{2}\hat{c}^{\dagger}_{n+1}U_{od}\hat{c}_{n}+h.c.,
\end{eqnarray}
with $\hat{c}_n$ is a vector of annihilation operators $\hat{c}_{n,\sigma,s}$ on
plane $n$,
$\hat{c}_n=(\hat{c}_{n,+,+},\hat{c}_{n,+,-},\hat{c}_{n,-,+},\hat{c}_{n,-,-})_n$, and
\begin{eqnarray}
U_d&=&(\mu-\cos{k_y})\sigma_x s_0-\cos{k_z}\sigma_z s_0\nonumber\\
&&-\sin{k_y} \sigma_y s_y+(\mu_s-t_z\sin{k_z})\sigma_y s_z\,,\\
U_{od}&=&-\sigma_x s_0+i\sigma_y s_z\,.
\end{eqnarray}
We numerically diagonalize this Hamiltonian and show the four closest
doubly degenete energy bands above and below zero energy in Fig.\ref{fig4}, with
OBC along $x$ from (a) to (c), and periodic boundary condition from (d) to (f) as comparison.
Panels (a) and (b) are for the topologically nontrivial phases with $\nu_{NL}^+=\nu_{NL}^-=1$
and $\nu_{NL}^+=1,~\nu_{NL}^-=0$,
and the spectra show edge states with two or one Dirac cone, respectively.
Panel (c) is for the topologically trivial phase with $\nu_{NL}^+=\nu_{NL}^-=0$,
where there is no edge state connecting the conduction and valence bands.

\emph{A real material example.-}Finally, we apply our method to 3D topological insulators of the
Bi$_2$Te$_3$ family\cite{real}.
These materials possess time reversal symmetry and are characterized by a $Z_2$ number.
Nevertheless, their topological nature is determined by the physics near the
time-reversal-invariant point $\Gamma(0,0,0)$ in the Brillouin zone,
around which the low-energy effective Hamiltonian also satisfies a chiral symmetry.
This Hamiltonian is given by
\begin{eqnarray}
H(k)=\epsilon_0(k)I_{4\times4}+\left(
\begin{array}{cccc}
M(k) & A_1k_z & 0 & A_2k_-\\
A_1k_z &-M(k)  &A_2k_-  & 0\\
0 &A_2k_+  &M(k)  & -A_1k_z\\
A_2k_+  &0  &-A_1k_z  & -M(k)
\end{array}\right),
\end{eqnarray}
with $k_{\pm}=k_x\pm ik_y$, $\epsilon_0(k)=C+D_1k_z^2+D_2|k_{+}|^2$ and
$M(k)=M-B_1k_z^2-B_2|k_+|^2$.
Using Dirac matrices, this Hamiltonian can be written as
\begin{eqnarray}
H(k)=M(k)\sigma_z s_0+A_1k_z\sigma_x s_z+A_2k_x\sigma_x s_x+A_2k_y\sigma_x s_y,
\label{bi2te3}
\end{eqnarray}
where we left out the identity matrix as it only changes the shape of the energy bands,
not the topology of the system.
The chiral operator is given by the absent fifth Dirac matrix,
$S=\sigma_y s_0$.

$P*T$ symmetry is here implemented by $P=\sigma_z$. However, we note that
particle-hole symmetry also exists in this case, which reads
%$CHC^{-1}=-H$, with $C=\sigma_y s_yK$
$CH^*(-k)C^{-1}=-H(k)$, with $C=\sigma_y s_y$,
%where $K$ denotes complex conjugation,
%$P H(k)P^{-1}=-H(-k)$, with $P=\sigma_y s_y K$
and satisfies $CC^*=1$.
Thus the model Eq.(\ref{bi2te3}) falls into the DIII class,
which is also characterized by a $Z$ invariant in 3D\cite{classes}.
Similar to our previous discussion, we write  $H=H_1+H_2$, with
\begin{eqnarray}
H_{1}(k)&=&M(k)\sigma_z s_0+A_1k_z\sigma_x s_z\,,\\
H_{2}(k)&=&A_2k_x\sigma_x s_x+A_2k_y\sigma_x s_y\,.
\end{eqnarray}
Then $H_{1}$ has a NL in $k_x-k_y$ plane, and the nodes of $H_{2}$
produce a single line enclosed by the loop.
This gives a NL winding number $\nu_{NL}=1$,
which corresponds to the topological properties of the $\Gamma$ point.

\emph{Summary.-}
In summary,
%we have studied a minimal model of NL semimetals in 3D with anticommuting gap terms.
we have studied Hamiltonians with $P*T$ symmetry where two terms define a NL which is gapped
out by the other terms.
In the presence of chiral symmetry, these gap terms can drive the system into a chiral TI,
which can be described by a winding number defined along the NL.
This winding number is associated with the geometric relation between the NL
and the zeroes of the gap terms.
We investigated a  lattice model in detail, which has two NLs in the BZ,
each of them with  a winding number of $1$ or $0$
due to the gap terms. This winding number corresponds to the emergence of a Dirac
cone for the surface states under OBC.
%, thus the model has three topologically different phases with the total winding number of $0$, $1$ or $2$.
Finally, we applied our method to the 3D topological insulators of the Bi$_2$Te$_3$ family,
and showed the connection between their topological nature and the NL winding number.

\emph{Acknowledgments.-}
Financial support from FCT through grant UID/CTM/04540/2013 is acknowledged.
S. C. is supported by NSFC under Grants No. 11425419, No. 11374354 and No. 11174360.

\end{document}